
\magnification=1100
\baselineskip 0.9truecm
\vsize 20.0 truecm
\hsize 13.0 truecm
\hoffset 0.5 truecm
\def\skuno{\vskip 20pt}
\def\skdue{\vskip 50pt}
\def\ltsim{\raise 2pt \hbox {$<$} \kern-1.1em \lower 4pt \hbox {$\sim$}}
\def\gtsim{\raise 2pt \hbox {$>$} \kern-1.1em \lower 4pt \hbox {$\sim$}}
\tolerance 3000

\parindent 0pt
\centerline {\bf VLBI OBSERVATIONS OF LOW POWER RADIO GALAXIES}
\vskip 1.0truecm
\parindent 0pt
G. Giovannini$^{1,2}$, W.D. Cotton $^3$, L. Feretti$^{1,2}$, L. Lara$^{2,4}$,
T.Venturi$^2$, J.M. Marcaide$^5$
\skuno
\parindent 0pt
1) Dipartimento di Astronomia, Bologna, ITALY
\goodbreak
\parindent 0pt
2) Istituto di Radioastronomia, Bologna, ITALY
\goodbreak
\parindent 0pt
3) NRAO - Charlottesville - USA
\goodbreak
\parindent 0pt
4) Instituto de Astrofisica de Andalucia, C.S.I.C., SPAIN
\goodbreak
\parindent 0pt
5) Departamento de Astronomia de Valencia, SPAIN
\skuno
classification: Astronomy
\skdue
Invited paper at the National Academy of Sciences Colloquium: "Quasars and AGN:
High Resolution Radio Imaging" - Irvine (CA) March 24 and 25, 1995
\skuno
Proceedings of the National Academy of Sciences - USA  in press
\vfill
\eject
\baselineskip 0.5truecm
\vsize 21.0 truecm
\hsize 15.0 truecm
\hoffset 0.5 truecm
\def\skuno{\vskip 15pt}
\def\skdue{\vskip 30pt}
\def\ltsim{\raise 2pt \hbox {$<$} \kern-1.1em \lower 4pt \hbox {$\sim$}}
\def\gtsim{\raise 2pt \hbox {$>$} \kern-1.1em \lower 4pt \hbox {$\sim$}}
\tolerance 3000

{\bf Abstract}
\skuno
The parsec scale properties of low power radio galaxies are reviewed here,
using the available data on 12 FR I galaxies.
The most frequent radio structure is an asymmetric parsec-scale
morphology, i.e. core and one-sided jet.
It is shared by 9 (possibly 10) of the 12 mapped radio galaxies.
One (possibly 2) of the other galaxies has (have) a two-sided jet emission.
Two sources are known from published data to show a proper motion; we present
here evidence for proper motion in two more galaxies.
Therefore, in the present
sample we have 4 radio galaxies with a measured proper motion. One of these
has a very symmetric structure and therefore should be in the plane of the sky.
The results discussed here are in agreement with the predictions
of the unified scheme models. Moreover,
the present data indicate that the parsec scale structure in low
and high power radio galaxies is essentially the same.
\skdue
\centerline {\bf Introduction}
\skuno
The knowledge of the structure of radio galaxies on the parsec scale is
important in order to test current models of jet dynamics as well as
radio source unification schemes.
VLBI data on powerful radio galaxies and quasars show strong evidence
of relativistic jets and in many cases a proper motion with an apparent
superluminal velocity  has been found (see Zensus this volume; Vermeulen
this volume).
VLBI observations of low power radio galaxies are also necessary
to compare the parsec scale properties of radio
sources with different radio powers and to test the unified scheme models,
which predict that also low power radio galaxies should have parsec scale
jets moving at a velocity close to the speed of light.
In this paper we will present and discuss the available VLBI data on extended
low power radio galaxies (Fanaroff-Riley Type I, hereafter FR I, [1]).
We will use the radio galaxies of the sample currently under study by us [2]
observed at 5 GHz with the VLBA or the global array.
This sample consists of the B2 and 3CR galaxies having a core flux
density greater than 100 mJy at 6 cm at arcsecond resolution [2].
The core flux
limit, imposed by observational constraints, could produce a sample biased
toward objects with jets pointing toward the observer. This point is not
important in discussing single obiects but has to be taken in account for
statistical studies. The well known FR I galaxy NGC 6251 (see [3] and
references therein), not included in our sample, was added, for a total of 12
FR I radio galaxies.
Some of them have been
observed also at 1.6 and 8.4 GHz. Only 5 have observations at different epochs
to search for a possible proper motion.
\goodbreak
\parindent 20pt
Radio galaxies in the same range of total radio power, but unresolved on
the arcsecond
scale, have not been properly mapped yet. Therefore, they will
not be discussed here. Observations are in progress to study this class
of radio sources, in order to understand their nature and connection with the
more powerful CSO and CSS sources (see Readhead this volume).
\goodbreak
\parindent 20pt
A Hubble constant H$_0$ = 100 km sec$^{-1}$ Mpc $^{-1}$ and deceleration
parameter q$_0$ = 1 have been used throughout this paper.
\skdue
\centerline {\bf Radio Morphology}
\skuno
The list of radio galaxies studied so far and discussed here is presented in
Table 1.
A morphological analysis based on the available VLBI maps
indicates that an asymmetric morphology, i.e. core and one-sided jet, is the
most frequent radio structure (see for example Fig. 1a,b).
It is shared by 9 (possibly 10) out of the 12 mapped radio galaxies.
A clear symmetric structure is found in 3C338
(Fig. 2), while 3C272.1 shows a complex structure with a
possible counter-jet close to the core (Fig. 3).
The one-sided jet is always well collimated and only small
oscillations or bendings are visible.
The nuclear emission at 5Ghz is always the dominant component.
When maps at two or more frequencies are available, the core emission
shows a flat or inverted spectrum while the jet emission has a spectral
index \gtsim 0.5. A few peculiar sources are described in detail below.
\goodbreak
\parindent 0pt
{\bf 3C264} - The one-sided jet detected at parsec resolution (Fig. 1a)
is oriented within a few
degrees of the optical jet visible in the Hubble Space Telescope map [8].
Unfortunately, the resolution of the radio map is too high for a detailed
comparison with optical data.
\goodbreak
\parindent 0pt
{\bf 3C272.1} - The core radio power in this nearby galaxy is low with respect
to the total radio power, so we expect that this source is very close to
the plane
of the sky (see next section). The VLBI map shows a complex structure with
a curved jet on the side of the main kpc jet (North) and a possible short
counter-jet (Fig. 3) on the opposite side.
\goodbreak
\parindent 0pt
{\bf 3C274} - See Biretta (present volume) for a detailed discussion of
this source.
\goodbreak
\parindent 0pt
{\bf 1144+35} - This source shows at kpc scale a
dominant core and two slightly asymmetric faint jets [9].
The core is strongly variable. Several measurements of the flux density [2]
at 1.4 and 5 GHz show that it was about 300 mJy in 1974.0 while now
(1995.1) it is
540 mJy after reaching a maximum of 610 mJy in 1991. Simultaneous
multifrequency observations
show that the core spectrum is flat between 1.4 and 5 GHz but
strongly steepens between 5 and 8.4 GHz.
The VLBI structure consists of two main components (A and C)
with an inverted spectrum between 1.7 and 5 GHz and low brightness, jet-like
features departing from them (Fig. 4). A comparison between our data and the
5 GHz VLBI map
obtained in the second Caltech-Jodrell Bank VLBI survey [10] shows that:
a) the component A is probably variable, therefore we tentatively identify it
as the core; in this case the parsec scale structure would be in the direction
of the fainter kpc scale jet. However, owing to the complexity of this source
and the slight asymmetry of the faint kpc scale jets, the definition of a main
jet may be ambiguous;
b) the separation between A and C increases between the two observing
epochs.
This proper motion is clearly visible also comparing the model given by [10]
with our visibilities. The data are consistent with a proper
motion of component C with respect to A with an apparent superluminal
velocity = 1.2c ($H_0=100$).
The snap-shot data obtained by us [2] are in agreement with this motion.
\goodbreak
\parindent 0pt
{\bf 3C338} - This source has a very steep global spectrum and is
associated with the
multiple-nuclei cD galaxy NGC6166. Even for this source the arcsecond core flux
density is strongly variable in time.
At pc resolution, this source shows a flat spectrum core with two
symmetric jets oriented in the E-W direction (Fig 2). We observed this source
at 1.6, 5 and 8.4 GHz, and
second epoch observations are available at 5 and 8.4 GHz. While the
analysis of the 5 GHz data
is still in progress, a preliminary comparison between the 8.4 GHz maps
of the two epochs (Fig. 5) shows a clear change in the source structure.
It is not obvious in such a complex structure how to determine an unambiguous
proper motion; however the present data suggest a possible motion
corresponding to an apparent velocity of 0.5 c.
\goodbreak
\parindent 0pt
{\bf 3C465} - This giant Wide Angle Tail radio galaxy has a total radio power
intermediate between low and high power radio galaxies. The parsec scale map
(Fig. 1b) shows a core emission and an asymmetric jet in the same direction
of the main kpc jet. A faint counter jet is visible only in VLA maps at
arcsecond resolution.
\skdue
\centerline {\bf Discussion}
\skuno
A - {\it Jet velocity and orientation}
\goodbreak
\parindent 0pt
In all the sources discussed here the parsec
scale jet is oriented on the same side of the main kpc scale jet with the
exception of 1144+35 whose interpretation is still uncertain (see before).
This correlation implies either that jets are intrinsically
asymmetric or that parsec and kpc scale jets are both relativistic.
The presence of relativistic jets in strong radio sources is now widely
accepted ([11] and Zensus present volume).
Furthermore, a detailed study of the inner kpc scale properties
of low power radio galaxies (Laing present volume) and the evidence of
proper motion at high velocities in some galaxies (see below)
suggest that radio jets in FR I  radio galaxies are initially
relativistic.
For these reasons, we interpret the radio structures presented here as
affected by Doppler
favoritism and will use the available data to constrain the possible values
of the jet velocity ($\beta$ = v/c) and of the orientation of the radio source
with respect to the line of sight ($\theta$).
We can constrain these two parameters in a four different ways:
a) from the jet to counter-jet brightness ratio, b) from the prominence of the
core radio power with respect to the total radio power, c) from comparing the
observed X-Ray nuclear emission with that expected
by the Self Compton Model, d) from imposing an upper limit on
$\theta$ to restrict the maximum intrinsic radio source size to 1.5 Mpc.
A detailed discussion on these methods can be found in [5].
The allowed values for the jet velocity $\beta$ and its orientation with
respect to the line of sight $\theta$ are given in Table 2.
\goodbreak
\parindent 0pt
B - {\it Proper motion}
\goodbreak
\parindent 0pt
While a proper motion of well defined features inside the radio jets is firmly
established and well studied in strong radio galaxies, quasars and BL-Lac
type objects (Vermeulen this volume; Zensus this volume), the situation is
still unclear for low power radio galaxies. A few galaxies (5) among
those presented here have at least two observations at different epochs
that can be used to look for the existence of a proper motion.
\goodbreak
\parindent 20pt
The galaxy 3C274 shows evidence of stationary knots as well as structures
moving at a sub-relativistic velocity; moreover some sub-structures seem to
move
with an apparent velocity larger than c (Biretta this volume).
NGC6251 could have both a stationary knot and one moving at v = 1.2c, the same
velocity we have found for the component C in 1144+35. 3C338 has certainly
changed its structure and a proper motion with v $\sim$ 0.5c is compatible with
data but it needs to be confirmed. For NGC315 an upper limit of 0.5c on the
jet velocity was derived in [4].
\goodbreak
\parindent 20pt
We appear to see stationary as well as subluminal and superluminal
knots in the jets of low power radio galaxies.
This could reflect the presence of oblique shocks and complex
situations where the measured velocity could be much lower than the
jet velocity (Begelman this volume).
More data are therefore necessary to  properly discuss this point, but in any
case the detection of proper motion in some low power radio galaxies
confirms that also in this class of radio galaxies parsec scale jets are
relativistic.
\goodbreak
\parindent 0pt
C - {\it Unified Models}
\goodbreak
\parindent 0pt
The present data and the derived values for $\beta$ and $\theta$ are in
agreement with the expectations from unified models. In fact in all the
low power
radio galaxies presented here, observational data are in agreement with the
presence of a parsec scale jet with a Lorentz factor $\gamma$ \gtsim 3,
viewed at angles larger than 30$^o$ with respect to the line of sight.
This is consistent with FR I sources being the parent population of BL-Lac
type objects.
\goodbreak
\parindent 20pt
Moreover we note that the parsec scale properties of FR I radiogalaxies are
very
similar to the parsec scale properties of FR II radio galaxies and quasars.
The large morphological difference between FR I and FR II radio galaxies at
the kpc-scale does not exist at the pc scale.
This similarity suggests that the nature and the power of the nuclear engine
is the same in low power and high power sources. The kpc scale differences
seem to arise from conditions far from the nuclei and could be related to a
different interaction with the surrounding medium.
A similar result was deduced by De Young [12] in an optical study of FR II
and FR I galaxies
and by Maraschi and Rovetti [13] who compared BL-Lac type objects and flat
spectrum radio quasars.
\skdue
\centerline {\bf Acknowledgments}
\skuno
We thank the staffs at the telescopes for their contribution to these
observations and the staffs at the Bonn and VLBA correlator where the data
have been correlated absentee. The National Radio Astronomy Observatory is
operated by Associated Universities, Inc., under contract with the National
Science Foundation.
\skdue
\centerline {\bf References}
\skuno
\parindent 0pt
1. Fanaroff, B.L., Riley, J.M. (1974) $\underline {M.N.R.A.S.}$ 167, 31
\goodbreak
\parindent 0pt
2. Giovannini, G., Feretti, L., Comoretto, G. (1990) $\underline {ApJ}$ 358,
159
\goodbreak
\parindent 0pt
3. Jones, D.L., Wehrle, A.E. (1994) $\underline {ApJ}$ 427, 221
\goodbreak
\parindent 0pt
4. Venturi, T., Giovannini, G., Feretti, L., Comoretto, G., Wehrle, A.E.
\goodbreak
\parindent 20pt
(1994) $\underline {ApJ}$ 435, 116
\goodbreak
\parindent 0pt
5. Giovannini, G., Feretti, L., Venturi, T., Lara, L., Marcaide, J., Rioja, M.,
\goodbreak
\parindent 20pt
Spangler, S.R., Wehrle, A.E. (1994) $\underline {ApJ}$ 435, 116
\goodbreak
\parindent 0pt
6. Venturi, T., Castaldini C., Cotton, W.D., Feretti, L., Giovannini, G.,
\goodbreak
\parindent 20pt
Lara, L., Marcaide J.M., Wehrle, A.E. (1995) $\underline {ApJ}$ submitted to
\goodbreak
\parindent 0pt
7. Feretti, L., Comoretto, G., Giovannini, G., Venturi T., Wehrle, A.E.
\goodbreak
\parindent 20pt
(1993) $\underline {ApJ}$ 408, 446
\goodbreak
\parindent 0pt
8. Sparks W.B., Biretta, J.A., Macchetto, F. (1994) $\underline {ApJSS}$ 90,
909
\goodbreak
\parindent 0pt
9. Parma, P., de Ruiter, H.R., Fanti, C., Fanti, R. (1986) $\underline {Astron.
A. S.}$ 64, 135
\goodbreak
\parindent 0pt
10. Henstock , D.R., Browne, I.W.A., Wilkinson, P.N., Taylor, G.B.,
\goodbreak
\parindent 20pt
Vermeulen,
R.C., Pearson, T.J., Readhead, A.C.S. (1995) $\underline {ApJSS}$ in press
\goodbreak
\parindent 0pt
11. Antonucci, R.R.J. (1993) $\underline {A.R.A.A.}$ 31, 473
\goodbreak
\parindent 0pt
12. De Young, D.S. (1993) $\underline {ApJ}$ 405, L13
\goodbreak
\parindent 0pt
13. Maraschi, L., Rovetti, F. (1994) $\underline {ApJ}$ 436, 79
\goodbreak
\parindent 0pt
\vfill
\eject
\baselineskip 15pt
\vsize 24.5 truecm
\hsize 15.0 truecm
\nopagenumbers
\parindent 0 pt
\skdue
{\bf Table 1 - VLBI Radio Galaxies}
\skuno
\settabs 8\columns

\+ Name &  & z  & Log P$_{408}$ & VLBI  & & Reference & & & & & \cr
\+      &  &    &    W/Hz       & morphology & & & & & & & \cr
\skuno
\+ 0055+30 & NGC 315 & 0.0167 & 23.95 & One-sided & & [4] & & \cr
\+ 0104+32 & 3C31    & 0.0169 & 24.50 & One-sided & & present paper & & \cr
\+ 0206+35 & 4C35.03  & 0.0375 & 24.28 & One-sided &  & present paper & & \cr
\+ 0755+37 & NGC2484 & 0.0413 & 25.04 & One-sided & & [5] & & \cr
\+ 0836+29 & 4C29.30 & 0.0790 & 25.08 & One-sided & & [6] & & \cr
\+ 1142+20 & 3C264   & 0.0206 & 24.85 & One-sided & & present paper & & \cr
\+ 1144+35 &         & 0.0630 & 24.15 & One-sided? & & present paper & & \cr
\+ 1222+13 & 3C272.1 & 0.0037 & 23.27 & Two-sided? & & present paper & & \cr
\+ 1228+12 & 3C274   & 0.0037 & 25.07 & One-sided & & Biretta present volume &
& \cr
\+ 1626+39 & 3C338   & 0.0303 & 25.25 & Two-sided & & [7] and present paper & &
\cr
\+ 1637+82 & NGC6251 & 0.0230 & 24.55 & One-sided & & [3] & & \cr
\+ 2335+26 & 3C465   & 0.0301 & 25.39 & One-sided & & [6] & & \cr
\skuno
Columns 1,2: radio galaxy names; Col. 3: total radio power at 408 MHz;
Col. 4: parsec scale radio morphology; Col. 5: reference for the radio
morphology.
\vfill
\eject
\vsize 24.5 truecm
\hsize 15.0 truecm
\nopagenumbers
\parindent 0 pt
\skuno
{\bf Table 2 - Jet velocity and orientation}
\skuno
\settabs 9\columns
\+ Name &  & & & allowed $\beta$ & & allowed $\theta$ & & $\theta$ if $\gamma$
\gtsim 3 & \cr
\+      &  & & &                 & &~~~~~$^o$          & &~~~~~$^o$  & \cr
\skuno
\+ 0055+30 & NGC 315 & & & 0.7 - 1.0 & & 30 - 40 & & 30 - 40 & \cr
\+ 0104+32 & 3C31    & & & 0.5 - 1.0 & & 25 - 60 & & 50 - 60 & \cr
\+ 0206+35 & 4C35.03 & & & 0.6 - 1.0 & &  0 - 50 & & 35 - 50 & \cr
\+ 0755+37 & NGC2484 & & & 0.5 - 1.0 & &  0 - 55 & & 40 - 55 & \cr
\+ 0836+29 & 4C29.30 & & & 0.7 - 1.0 & &  0 - 40 & & 30 - 40 & \cr
\+ 1142+20 & 3C264   & & & 0.5 - 1.0 & &  0 - 55 & & 40 - 55 & \cr
\+ 1144+35 &         & & & 0.4 - 1.0 & &  0 - 60 & & 45 - 60 & \cr
\+ 1222+13 & 3C272.1 & & & 0.4 - 1.0 & & 60 - 80 & & 75 - 80 & \cr
\+ 1626+39 & 3C338   & & & 0.0 - 1.0 & &  0 - 90 & & 85 - 90 & \cr
\+ 1637+82 & NGC6251 & & & 0.9 - 1.0 & & 40 - 45 & & 40 - 45 & \cr
\+ 2335+26 & 3C465   & & & 0.6 - 1.0 & &  0 - 50 & & 35 - 50 & \cr

\skuno
Columns 1,2: radio galaxy names; Col. 3: allowed range for $\beta$ = v/c
in parsec scale jets; Col. 4: allowed range for the source angle in degree
with respect
to the line of sight $\theta$ corresponding to the jet velocity range given
in Col. 3; Col. 5: as Col. 4 but for a jet with a
Lorentz factor $\gamma$ \gtsim 3.
\vfill
\eject
{\bf Figure Captions}
\skuno
\parindent 0pt
{\bf Fig. 1} - a) VLBI map of 3C264 at 5.0 GHz. The HPBW is 3.5 $\times$ 2.1
mas
in PA 11$^o$. The peak flux is 135 mJy/beam; contour levels are:
-0.3 0.3 0.5 0.8 1.5 3 5 10 25 50 100 mJy/beam. The arrow shows the direction
of the main kpc scale jet.
\goodbreak
\parindent 0pt
b) VLBI map of 3C465 at 8.4 GHz. The HPBW is 2.52 $\times$ 0.83 mas in
PA -9.7$^o$. The peak flux is 132 mJy/beam; contour levels are: -0.75 0.75
1.5 2 3 5 10 20 50 100 mJy/beam. The arrow shows the direction
of the main kpc scale jet.
\skuno
\parindent 0pt
{\bf Fig. 2} - VLBA map of 3C338 at 5 GHz. The HPBW is 2.2 $\times$ 2.2 mas.
The peak flux is 44.4 mJy/beam; contour levels are: -0.5 0.5 0.7 1 1.5 2 3 4 6
8 10 20 30 40 mJy/beam. The two-sided arrow shows the direction of the
symmetric kpc scale jet.
\skuno
\parindent 0pt
{\bf Fig. 3} - VLBI map of 3C272.1 at 5 GHz. The HPBW is 5.3 $\times$ 1.2 mas
in PA -5$^o$. The peak flux is 168 mJy/beam; contour level are: -0.7 0.7 1
2 3 5 7 10 20 50 100 150 mJy/beam. The arrow shows the direction of the main
kpc scale jet.
\skuno
\parindent 0pt
{\bf Fig. 4} - VLBI map of 1144+35 at 5.0 GHz. The HPBW is 2.5 $\times$ 0.8
mas in PA 5$^o$. The peak flux is 255 mJy/beam; contour levels are: -1.5 1.5
3 5 7 10 15 20 50 100 200 mJy/beam. The arrow shows the direction of the main
kpc scale jet (but see text).
\skuno
\parindent 0pt
{\bf Fig. 5} - {\it upper}: VLBI map at 8.4 GHz of 3C338 obtained on 1991.3.
The peak flux is 62.7 mJy/beam; contour levels are: -0.3 0.3 0.5 0.7 1 1.5 2 5
10 40 60 mJy/beam;
\goodbreak
\parindent 0pt
{\it bottom} VLBA map
of 3C338 at 8.4 GHz obtained on 1994.92. The peak flux is 25.8 mJy/beam;
contour
levels are: -0.3 0.3 0.5 0.7 1 1.5 2 5 10 15 20 mJy/beam.The HPBW is
2 $\times$ 1 mas in PA 0$^o$ in both maps which have been plotted in the same
scale.
\bye